\documentclass{elsart3-1}

\usepackage{graphicx}

\usepackage{amssymb}

\usepackage[english,francais]{babel}

\newtheorem{e-proposition}[theorem]{Proposition}

\newtheorem{e-definition}[theorem]{Definition\rm}

\renewcommand{\theequation}{\arabic{equation}}
\def\og{\leavevmode\raise.3ex\hbox{$\scriptscriptstyle\langle\!\langle$~}}
\def\fg{\leavevmode\raise.3ex\hbox{~$\!\scriptscriptstyle\,\rangle\!\rangle$}}

\begin{document}
\begin{frontmatter}


\selectlanguage{english}
\title{CP, T and fundamental interactions}


\selectlanguage{english}
\author[authorlabel1]{Jean-Marie Fr\`{e}re}, \ead{frere@ulb.ac.be}

\address[authorlabel1]{Service de Physique Th\'{e}orique, Universit\'{e} Libre de Bruxelles,
   Bvd du Triomphe,    B-1050 Bruxelles, Belgique }

\begin{abstract}
We discuss the importance of the CP (simultaneous particle-antiparticle and left-right
permutation) and T (time reversal) symmetries in the context of fundamental interactions. We show that they may provide clues to go beyond the 4-D gauge interactions. We insist on the fact that T violation is not associated to a
degradation (like in entropy), but simply characterized by different trajectories. \\

\vskip 0.5\baselineskip


\end{abstract}
\end{frontmatter}

\selectlanguage{english}
\section{Introduction}
\label{intro}
Why are CP (and T) symmetries important? The main purpose of this short
contribution is to try and answer this question.

There are on one side somewhat psychological reasons: on one hand, most of Nature's
laws are symmetrical under left and right (P) permutation (on the ``macroscopicÝÝ
level, handedness seems to be linked to living organisms chemistry or behaviour).
The same could be said of charge conjugation symmetry  (the exchange of particles and
antiparticles, which will be noted C )were it not for the fact that
antiparticles are virtually absent from our everyday observations (ignoring $\beta +$
decays and reactor antineutrinos !).
In a similar way, we are used to idealize mechanics and electromagnetism as
time-reversal invariant, and to blame any lack of such invariance on
entropy considerations (keeping in mind that such considerations have to be added "by hand"
in the formalism, and cannot be derived directly from the local interactions).

More important are somewhat more theoretical reasons, linked to gauge theories.
In this context, it can be affirmed that CP (or T) are
THE natural symmetry of gauge interactions in a mimimal Lagrangian
(that is, in absence of scalars interactions - including   fermion masses ).
This is precisely the importance of studying CP and T symmetries: their breaking
implies the presence \emph{something more} than the part of the Standard Model
of Fundamental Interactions that we know best, namely gauge theories.
At the very least, CP violation probes directly the still unobserved scalar
sector of the Standard Model.

One more reason for the central r\^{o}le of CP violation is linked to the matter-antimatter
asymmetry of the Universe. Here we are dealing with factual evidence, and the choice to
accept an elaborately tuned (specially in an inflation context) cosmological asymmetry,
or new sources of CP violation.

Before closing this introduction, it is also worth to mention the various T violating
processes, and to wonder if any relation exists between those different "arrows of time".
In this context, we must mention the increase of entropy (and its associated
impact on chemistry, life and death processes,
but also in the context of the construction of the matter-antimatter asymmetry),
the cosmological time (and the corresponding expansion of the Universe ),
and of course our present subject, microscopic time.

\section{The evidence for CP (or T) violation"}
From a pedagogical point of view, it is often difficult to single out the evidence for CP (and \emph{a fortiori} for T) violation. We will try to produce below
2 clear-cut examples.

\subsection{CP violation}

The simplest presentation to my knowledge involves the production of $e^+$ vs $e^-$
some distance away from a hadronic collision. To speak roughly, it can be formulated as follows: "arrange for a collision between 2 energetic particles \emph{or antiparticles}, put shielding around it, and move back a sufficient distance,
count the emerging $e^+ \pi^-$ vs $e^- \pi^+$ : there will always be a small excess of the $e^+$".
The important point is that the observed excess is always in the same direction (more $e^+$), whether we start from
colliding particles, or antiparticles.
This process is a clear violation of C,
while the integration over all directions ensures that the symmetry is not restored by P,
hence establishing the CP violation.

Of course the most evident set-up for the experiment
(dumping a beam of particles on a target made of ordinary matter) in itself violates
CP from the start (and thus a more detailed examination of
the process is then needed to establish the effect),
however a perfectly equivalent experiment can be made through proton-antiproton collisions. This set-up (similar to the CP-Lear experiment, which we evoke later) is symmetrical under CP and gives the same clear conclusion: particles can be distinguished from antiparticles in absolute terms, independently of the notion of left or right.

Although this subject is of course an illustration of the very classical $K^0
\ -\ \overline{K^0}$ system, delving in some more detail will help clarify some important issues, and
explain why the outcome is the same wether the initial state is a between  particle-particle, particle-antiparticule,
or (still a gedanken experiment) antiparticle-antiparticle

The mechanism is indeed familiar: the collision produces amongst other things neutral $K^0$
and $\overline{K^0}$ (in equal quantities) which survive long enough to emerge from the shielding. While the $K^0 \ - \ \overline{K^0}$ are eigenstates of gauge interactions,
they do mix through weak interactions. The "mass eigenstates" for free propagation are
then the "long-lived" and "short-lived" $K^0_L \ K^0_S$. Because CP violation is only
a small effect (parametrized here by $\epsilon$), it is convenient to re-write
these states in terms of the "CP" even and odd eigenstates $K^0_1 , K^0_2$, which,
modulo a suitable choice of phase conventions, read:

\begin{eqnarray} \label{Kdef}
 | K^0_1> &=& \frac{|K^0> + |\overline{K^0}>}{\sqrt{2}} \\
    |K^0_2> &=& \frac{|K^0> - |\overline{K^0}>}{\sqrt{2}} \\
  |K_L> &=& {1 \over \sqrt{1 + |\epsilon|^2}} (|K_2^0> + \epsilon |K_1^0>) \\
  |K_S> &=& {1 \over \sqrt{1 + |\epsilon|^2}} (|K_1^0> + \epsilon |K_2^0>) \\
\end{eqnarray}

These equations were only introduced here to show that the $K^0_L$ state, for instance,
proportions of $K^0$
and $\overline{K^0}$ differ by small amounts controlled by $\epsilon$, the usual
CP parameter more familiarly related to the ratio of 2 and 3-pion decays of the $K_L$.
Since the semi-leptonic decays are respectively ($K^0$ is
for historical reasons defined as the $ \overline{s} \gamma_5 d$ state)
$K^0 \rightarrow  e^+ \nu \pi^-$ and  $\overline{K^0} \rightarrow  e^- \overline{\nu} \pi^+$,
this results in the mentioned excess of the $e^+$ channel, when we look only at the $K_L$ decays
(i.e. "at a sufficient distance from the production" in the above recipe). The resulting
asymmetry ($\frac{e^+ - e^-}{e^+ + e^-}$, in those conditions is $ 2 Re(\epsilon)$, somewhat
less than 0.5 \%).

This is a good opportunity to discuss the relation between the manifestation of CP violation
and the observation of time dependence. Indeed, although no explicit time dependence was mentioned in
the discussion above, an implicit one arises from
the distance between the production and the observation point.  At first sight indeed the claimed excess of $e^+$ from $K^0_L$ decay at long distance could be compensated by the opposite effect
for the shorter-lived $K_S$ close to the production. There is some truth in this
indeed: at the very production site, we can just consider the $K$ and $\overline{K}$,
which produce equal quantities of $e^+$ and $e^-$ -channel decays. However the
total effect subsists even when one integrates over time (or distance) from the interaction point. This is
simply due to the fact that the $K_S$  decays much more rapidly in the $2 \pi$
channel, hence giving the semi-leptonic mode a smaller branching ratio than in $K_L$
(namely the branching ratio of $K_S \rightarrow \pi^\pm e^\mp  \stackrel{(-)}{\nu}$ is only $(6.9 \pm 0.4) 10^{-4}$,
while the corresponding decay probability for the $K_L$ is $(39.81\pm0.27) \%$.

 For this reason, the cancellation does not hold,
 and the very explicit CP asymmetry advocated here is observed without having to take into account
 the time dependence. We should also stress that this happy circumstance is due to an exceptional kinematical
 accident, namely the strong lifetime difference between the $K_L$ and $K_S$;  in most other cases,
  timing information will be needed to exhibit and measure CP violation; this is particularly true in the $B$ system.

We anticipate on the next section, which will show the r\^{o}le of phases to stress
already here that the CP effects are linked to unremovable phases involving more
than 2 states (here, the $K$, $\overline{K}$ and the final states, all in communication).

\subsection{T violation}
The issue of T violation is indeed more complex to show. For a long time,
T violation was only surmised from the fact that local Lagrangian theories
obey the TCP theorem, and hence a violation of CP implies a violation
of T in that context.

The obvious difficulty is that the T symmetry involves not only the reversal
of t as a kinematical variable (reversing all speeds and angular momenta,
for instance), but also and foremost a permutation of the initial and final states.
This makes the test very difficult, and in fact impractical in the case of
decays (or collisions). It is impossible indeed in practice to reconstitute
the kinematical (and phase) conditions of the final particles to make them
re-assemble in the original unstable one.

The situation is not completely desperate however, if we can turn to 1-particle
evolution.
We present 2 cases: one is currently observable, through tagging, while the
other, waiting for the hypothetical observation of an electric dipole moment
is for the time being a "gedanken" (but more spectacular) experiment.

We will once again take the example of the CP-Lear experiment \cite{Angelopoulos:1998dv}
(but many B physics
situations are similar). In the strong interaction production of neutral kaons,
it is possible to "tag" separately $K^0$ and $\overline{K^0}$ through the
accompanying particles:  simultaneous production of $K^-  K^0$ or $K^+  \overline{K^0}$,
 making use of the flavour-diagonal character of strong interactions.
At the moment of
decay, we have already indicated that the semi-leptonic final staate $\pi l \overline{\nu}$
allows for a similar tagging.

From this set-up, it is possible to compare
the probabilities $|<\overline{K^0}|S|K^0>|^2$ and $|<K^0|S|\overline{K^0}>|^2$,
and check that they are indeed different (remember that other states are accessible,
and will compensate those differences if needed -- see below CP vs CPT).
In passing, let us remark that the above transitions are NOT related by CPT,
which for instance relates $<\overline{K^0}|S|K^0>$  just to ...itself.

\bigskip

The above example relies of course on the reliability of the tagging (for instance,
that no explicit CP violation takes place in the decay vertex, at least to
the accuracy requested to establish the effect), and is still in some way indirect.

\begin{center}
\begin{figure}[h]
 \centerline{ \includegraphics[width=7cm]{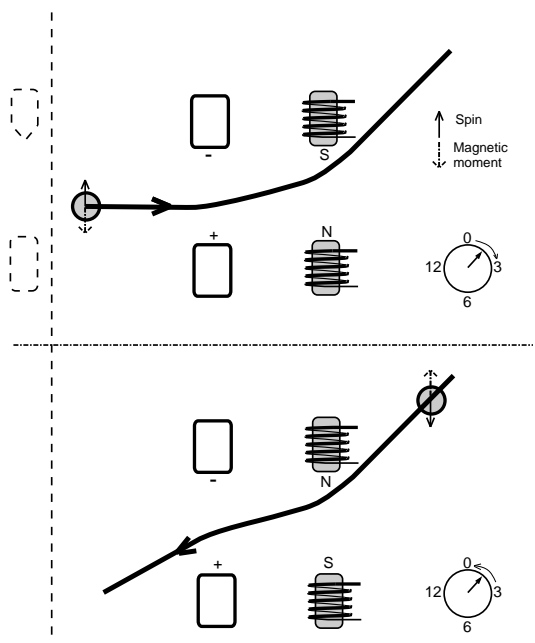}}
  \caption{Gedanken experiment: microscopic irreversibility in
  presence of an hypothetical electric dipole moment. The time-reversed trajectory
  is different, allowing to determine which is the "normal" and which is
  the "time-reversed" situation }\label{GedankenT}
\end{figure}
\end{center}

We propose below a more spectacular, but still speculative situation. Once again
it relies on the evolution of one single particle (here, the neutron). Let us assume
that the \emph{electric dipole moment} of the neutron has been measured. Such
a dipole moment is expected at a very small level in the Standard Model (of the order
of $ 10^{-32} e.cm$), much
below the current limit  $< 0.29 10^{-29} e.cm$ .This bound in fact already puts
severe constraints on most extensions "beyond" the Standard Model (Left-Right symmetrical
model, supersymmetry, ...).
The fact that an electric dipole moment for an "elementary" particle (or for the
ground state of an assembly of quarks, like the neutron) violates CP or T is a
bit tricky. The main point is that the only direction in which the dipole moment
can point (for such a particle at rest) is along the spin.
\begin{equation}\label{edm}
    \overrightarrow{d_n} = \kappa \overrightarrow{s_n}
\end{equation}
with $\kappa$ a fixed, calculable (measurable) coefficient;
(the situation is different, say
for a water molecule, where other directions exist and the large dipole moment
of course does not violate CP; it is also different in non-commutative theories,
where extra tensors are available). The relation (\ref{edm}) indeed violates P and T
because the first quantity $d_n$ is an "ordinary vector" (think of the distance
between the opposite charges of a dipole), it flips under P and is time invariant,
while the second (as an angular momentum), behaves in the opposite way.

We have reproduced a thought experiment (see Fig \ref{GedankenT} where such a neutron (with observable
electric dipole moment) is sent through a 2-phase "Stern-Gerlach" set-up,
where one stage uses an inhomogeneous magnetic field, and the other an electric one.
As seen from this (gedanken) drawing, the "time-reversed" trajectory is just different
from the initial one. (Note that the magnetic fields, linked to currents have
been reversed, but not the electric ones).

We produced this example to illustrate one point: in this "microscopic breaking
of Time invariance", no entropy comes into play: the return trajectory is in
no way "degraded" with respect to the first one - it is simply \emph{different}.
Just as someone taking one path to go to work in the morning, and systematically
a different one to return home!

\section{CP or T, the natural symmetries of gauge interactions ... in 3+1 dimensions}
In a way this is stating the obvious, but is probably not useless. We are indeed used to see the CP violation associated to a phase in the Cabibbo-Kobayashi-Maskawa matrix, which is closely associated to the exchange of the charged gauge bosons $W$. This may
lead to wrongly associating CP to such exchanges. Of course, there is no real paradox,
and the impression arises from the fact that, producing particles with accelerators,
we work spontaneously in the fermion mass-eigenstate basis.

Let us thus go back to basics. In 3+1 dimensions,
 the simplest fermion transforming consistently
 under the "proper" Lorentz group (rotations, boosts,
  without inclusion the reflections of the spatial or
  temporal coordinates) is represented by a spinor with 2 degrees of freedom.
 This  can consist in   a 2-component spinor field, i.e;, 2 complex numbers
  (fields) (technically it is called a semi-spinor of first or second species),
  or in a reduction of the more familiar 4-component "Dirac spinor", either by Weyl projection
  (on so-called Left or Right-handed modes), or by imposing a Majorana-like condition
  (the equivalence of these reductions is not a general feature when working
  in a different number of dimensions).

  We use below the Weyl representation, and take as example a "Left" projection.
  Assuming a massless fermion, the solutions of the Dirac equation can be developped
  as usual in plane waves, with the operators $a$ describing the destruction of
  a particle. For simlicity, we have used a sum over the various modes (an integral
  over the continuous variables is understood). As seen below, the sum includes
  both positive and negative-energy states, and the helicity ($\lambda$ is the projection
  of the spin on the direction of motion) is opposite to the sign of the energy,
  namely the $\Psi_L$ field describes a positive energy particle of negative helicity (L)
  and a negative energy particle of positive helicity (R).

\begin{equation}\label{fermion}
    \Psi_L (x,t) = \sum_{\scriptstyle p^0 > 0 \atop \scriptstyle  \overrightarrow{p}} a_{p^0,
\overrightarrow{p}, \lambda_-} {e^{-ipx} \over \sqrt{2 \omega}} u(\omega,  \overrightarrow{p},
\lambda_-) + \sum_{\scriptstyle p^0 < 0 \atop \scriptstyle  \overrightarrow{p}} a_{p^0,
\overrightarrow{p}, \lambda_+} {e^{-ipx} \over \sqrt{2 \omega}} u(-\omega,  \overrightarrow{p},
\lambda_+)
\end{equation}

We make the next step explicitly to avoid confusion between C and CP operations.
To avoid negative-energy particles, we proceed along the usual "trick" of replacing
the \emph{destruction} operator of the negative-energy mode (energy $- \omega$), $a_{-\omega,
\overrightarrow{p}, \lambda_+}$ by the \emph{creation} operator $b_{*\omega,
-\overrightarrow{p}, \lambda_+}^+$. In this operation, we are forced (to conserve
energy-momentum and angular momentum) to flip the 4 components of the 4-momentum,
\emph{and }the spin;  as a result, the helicity $\lambda_+$ (product of spin by 3-momentum) is
not affected in the process.. .For the same reason, other conserved quantities, like
the fermion number or electric charge are also flipped. The new state introduced is thus an antiparticle with Right helicity,
namely the CP (\emph{and not the yet inexistant C conjugate}) of the initial fermion.
Summation of dummy variables hides the process somewhat, but we finally reach:
\begin{equation}\label{FermionAnti}
    \Psi_L  (x,t) = \sum_{p^0 = \omega,  \overrightarrow{p}} \{ a_{\omega, \overrightarrow{p},
     \lambda_-}
{e^{-ipx} \over \sqrt{2\omega}} u(\omega, \overrightarrow{p}, \lambda_-) +
 b_{\omega, \overrightarrow{p}, \lambda_+}^+
{e^{ipx} \over \sqrt{2\omega}} v(\omega, \overrightarrow{p}, \lambda_+)\}
\end{equation}

With this result in hand, it is straightforward to check that gauge interactions
between fermions automatically respect CP (remember, we have not introduced masses
yet). We will take the significant example of a charged gauge boson interacting
with two fermions:
\begin{equation}\label{gauge interaction}
   g \  W^{\mu} \overline{\psi_{Lu}} \gamma_{\mu} \psi_{Ld} +h.c.
\end{equation}
Once this equation is expanded in terms of the creation an destruction operators
(as in eq.(\ref{FermionAnti})), it is obvious that the second term is just the CP
conjugate of the first. CP invariance results since the coefficients of both terms
are identical (note that we did not need to introduce the charge or CP conjugation
matrix C: it is only needed if we want to face the case of Majorana masses -
a possibility in the leptonic sector, which will not be considered further here).

This is sufficient to claim a very important result, which is actually independent
of the number of spatial dimensions (this will be used later): gauge interactions
alone (i.e. in absence of scalar interactions, which include the mass terms)
respect CP as a symmetry. In 3+1 dimensions, neither C nor P are granted for
gauge interactions, but CP of course is.
Note that even the triangular anomalies don't take exception to this: in the
massless case, the effective T and CP violating term
$\theta \epsilon_{\mu \nu \rho \sigma} F^{\mu \nu} F^{\rho \sigma}$ can always
be rotated away.

\bigskip

We now remind very briefly how CP violation is introduced in the phenomenological
Cabibbo-Kobayashi-Maskawa way before evoking in the next section possible more
fundamental sources.

The above argument about CP invariance does not apply to the scalar couplings
(in practice, Yukawa couplings since raw mass terms are not allowed in the Standard
Model). Consider indeed the Yukawa couplings: ($\Phi_a$ are  Brout-Englert-Higgs scalar
doublets -- we allow for several of them, labeled by the index $a$ --, while $\Psi_{Lj}$ is a left-handed quark
doublet field associated to the fermion family $j$, and $u_{Ri}$ similarly describes the right-handed "up" quarks,
labeled by $i$ for (u,c,t) )

\begin{equation}\label{Yukawa}
    \lambda_{a ij} {\overline{u_{Ri}} }\Phi_a^+ \Psi_{Lj} +hc
\end{equation}

Here indeed, the matrix $\lambda $ is in general complex, and distinct
from its complex conjugate. Furthermore, relative phases between the vacuum
expectation values in case of more than one scalar fields can add
another source of CP violation.
(the latter case will be referred to as "spontaneous CP violation" if the
Lagrangian is otherwise CP invariant, and the only breaking originates in
the relative phases of vacuum expectation values).

After diagonalization of the mass matrix, the Left- and Right-handed charged quarks
currents in the new basis read:( $K^{L(R)}$ being the Kobayashi-Maskawa mixing matrix)
\begin{equation}\label{KM}
    j^{\mu}_{L,(R)} = {\overline{u}_{i L(R)}} \gamma^{\mu} K_{ij}^{L(R)} d_{j L(R)}
\end{equation}
In principle, each of these mixing matrices contains $n (n+1)/2$ phases, where $n$ is
the number of fermion families; however as is well-known, not all the phases are observable.
It is thus possible to redefine the left-handed fermion fields to reduce the number
of $K^L$ phases to $(n-1) (n-2) /2$, at the cost of a rotation of the right-handed fields
to keep the mass matrix real; therefore the count of phases in $K^R$ in general stays
maximal. It is only because the Standard Model only gauges the left-handed currents (hence
$K^R$ can be ignored) that the famous result that 3 families are needed for CP violation
holds. In the leptonic sector, the same count is valid, except in the case where neutrinos
are Majorana particles (the PMNS matrix): in that case, the right-handed fields cannot
be used to make the neutrino mass matrix real, and (n-1) additional phases must be included.

\section{Which object should we call "Antiparticle"}
Before moving to more speculative ground, we may pause for a question of nomenclature. What object should we call "antiparticle"?
As long as only quantum electrodynamics was involved, the definition did not matter much, since, as far as particles interact only with the photons, both the $C$ and $CP$ conjugates are present (for the$ e^-_L$, respectively the $e^+_L$ and the $e^+_R$). In the framework of the Standard Model however, the fields appearing don't respect the $C$ symmetry, as we have seen, and the  $e^-_L$ and $e^+_L$ belong to completely different fields, with different $SU(2)$ properties. For the neutrino, we don't even know if the $\nu_R$ state exists at all.
It may thus be expedient (and some other articles in this issue have quite independently taken the same view)
to call "antiparticle" the $CP$ conjugate state to the particle.
This does not imply any confusion, as one can keep the appellations "C-conjugate", "P-conjugate"
whenever one needs to refer specifically to these objects (nowadays much less frequently).

\section{Towards a fundamental origin of CP violation}
The  Kobayashi-Maskawa approach describes correctly the known CP violation,
and the result in this latter respect is quite impressive; still, this appears
more as a successful parametrization than a fundamental understanding.

The main difficulty with CP violation is that, if we look for a more fundamental
theory to avoid the Yukawa couplings (and ideally the arbitrary scalars), the
most logical choice is to turn to gauge interactions, which, as we have seen
find in CP their natural symmetry. This is precisely one of the reasons why
we consider that CP violation is a key to understanding physics beyond the standard
model, and a challenge which must be faced in any fundamental approach.

We would like to hint here at 2 possibilities. The first one takes place in the
usual 3+1-dimensional context, in so-called "dynamical symmetry breaking". In such
theories, the usual scalars are replaced by bound states of fermions, in a "pure gauge"
context
(bound for instance through a new extra-strong interaction, often referred to as "technicolor").
In such a case, the only solution is to have these effective scalars develop several
vacuum expectation values (condensates), whose relative phases then lead to "spontaneous
CP violation", as mentioned about eq. (\ref{Yukawa}).

\bigskip

One other interesting possibility is to have the CP violation originate from a pure
gauge theory in more than 3+1 dimensions.
The principle can be explained quite simply, by turning back to the gauge interaction,
but this time in 4 or 5 spatial dimensions. The spinors used are now (at least) 4-component ones,
and contain (as usual Dirac spinors) both the Left- and Right-handed fermions (in terms
of a 3+1 reduction):

\begin{equation}\label{GaugeExtaDim}
   \overline{ \psi} V_{M} \Gamma^M \psi
\end{equation}

We distinguish now between the usual 3+1 dimensions (called $\mu$ and the remaining ones:
$M=\mu, \ 4, \ 5$, and assume the remaining ones are  compactified (either on a torus with a radius $R$,
such that $1/R >>$ the accessible energy -- say several TeV, or on a compact orbifold).
It is then usefull to single out the inaccessible components, and one observes terms like
(again in 3+1-dimensions language)

\begin{equation}\label{RXtradim45}
    \overline{\psi}_L  ( i \gamma_5 W^4 + W^5) \psi_R
\end{equation}
Such terms correspond to pseudoscalar and scalar interactions respectively,
and can generate complex mass terms.
To be more accurate, we must keep in mind that the quantities above are in fact
gauge-dependent; however the integration over the extra dimensions takes care of this,
and bring in a new quantity in the equation, namely the  integral (or flux) of
the gauge field along the extra dimensions.

For instance in 4+1 dimensions, the "Hosotani loop" is definitely a way to induce CP violation
in an otherwise CP conserving gauge theory, through the term \cite{Cosme:2002zv}:

\begin{equation}\label{Hosotani}
   \overline{\psi}_L \oint ( i \gamma_5 W^4) dx^4\psi_R
\end{equation}

In this context, the Hosotani terms appear as some extra elements (similar to boundary
conditions, bringing extra parameters) in the theory, and can actually break both CP and the initial gauge group.

Introducing such an approach however requires to consider larger gauge groups. \cite{Cosme:2002zv}

\section{CP versus TCP, and the Matter-Antimatter asymmetry}
If we apply the TCP invariance to the special case of the survival
of a single particle $X$,
we see the matrix element is identical to that of the antiparticle $\overline{X}$

$$< X \mid S \mid X > = < \bar{X} \mid S \mid \bar {X} >$$

\noindent A comparison to the familiar expression :
$$< X \mid S \mid X > = e^{-i (m -i \Gamma /2)(t-t_0)} $$
establishes that particle and antiparticle have both equal masses
and equal "total decay width" (the inverse of the lifetime)  $\Gamma = 1/ \tau$
This equality of the lifetimes makes it impossible to imagine a cosmological
scheme where the initially equal numbers of particles and antiparticles produced
in connection with gravity decay at different speeds.

\bigskip

It still allows however for the creation of a matter-antimatter imbalance.
The reason is that TCP only constrains the total decay probabilities to be equal,
but not the partial ones.
In other terms, $X$ and $\overline{X}$ live the same time, but can suffer
different deaths...

More explicitly, let consider a particle $X$ with only the  2 decay
processes $X \rightarrow a , X \rightarrow b$, and the charge
conjugate processes, $\bar X \rightarrow \bar a , \bar X \rightarrow
\bar b$.
Let us adopt the notation:($f$ is any of the final states)
$$A_{X \rightarrow f} = < f \mid S \mid X >$$ for the amplitude, while we use $P$ for
the transition probability: $P_{X \rightarrow f} $.

Summing over all possible decay channels $f$, TCP implies
$$\sum_f {P_{X \rightarrow f }} =\sum_f {P_{\bar X \rightarrow \bar f} }$$
but \emph{\textbf{does not}} imply

$$P_{X \rightarrow a } = P_{\bar X \rightarrow \bar a}$$
as long as the difference is compensated by other decay channels!

\bigskip

In particular, if the channels $a$ and $b$ have different baryon (or lepton)
number, a net baryon (lepton number) is created.

It should be clear however that whatever mechanism allowing for a difference
between ${P_{X \rightarrow a }}$ and ${P_{\bar X \rightarrow \bar a} }$
must do so in such a way that the total lifetimes are kept equal, namely,
the calculation leading to this difference must in some way know of
the existence (and physical availability) of the other channels.In the calculation, this indeed typically occurs through higher-order contributions,
and in the present case, it requires that the $a$ final state can be reached
either directly, or by re-scattering $X \rightarrow b ; b\rightarrow a$. The fact
that the $b$ channel must be kinematically accessible is traduced in the following
graphical illustration by the presence of a non-vanishing "unitarity cut".

\begin{center}
\begin{figure}
\centerline{\includegraphics [width=7cm]{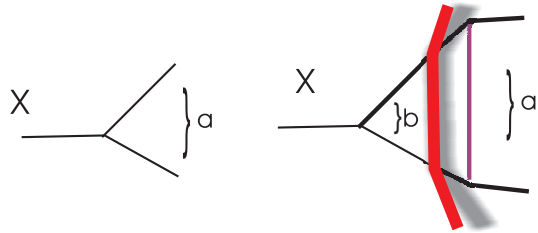}}
\caption{interference between channels a and b, responsible
for differences in the decay branching ratios of $X$ and $\overline{X}$} \label{unitarityCut}
\end{figure}
\end{center}

It is not our purpose to go here into the details of baryo- or lepto-genesis,
but it may be useful to notice that here again, a 3-state scheme comes into play,
allowing for physical (unremovable)  phases between the direct $X\rightarrow a$ and
the indirect $X \rightarrow b ; b\rightarrow a$, a situation we already found in
the Kaon system (direct $K^0$ decay, or via $K^0 - \overline{K^0}$ transition).

We just close by reminding that for a baryo-or leptogenesis scheme to be successful, new mechanisms for
CP violation must be found, as the usual Cabibbo-Kobayashi-Maskawa is notoriously insufficient
for this purpose.

\section*{Acknowledgements}
This work was supported in part by the French Community of Belgium (IISN),
and by the Belgian Federal Policy Office (IAP VI/11).




\section* {Remark}
Much of the material presented above is covered (usually from a different
perspective) in textbooks; we just mention a few extra references dealing
more precisely with the above approach.

\end{document}